\documentclass[aps,prx,reprint,longbibliography,amsmath,amssymb,nofootinbib,floatfix]{revtex4-2}
\usepackage{amsthm,graphicx,bm,booktabs,hyperref}
\usepackage[T1]{fontenc}
\hypersetup{hidelinks,pdftitle={The Cost of Quantum State Conversion},pdfauthor={Ansh Mishra and Aryan Senthilkumar}}
\graphicspath{{figures/}}
\newtheorem{theorem}{Theorem}
\newtheorem{lemma}{Lemma}
\newtheorem{proposition}{Proposition}

\newcommand{\B}{\mathcal B}
\newcommand{\C}{\mathcal C}
\newcommand{\tr}{\operatorname{tr}}

\newcommand{\id}{\mathbb I}
\newcommand{\eps}{\varepsilon}
\newcommand{\cost}{\mathsf S}
\newcommand{\cl}{\rightsquigarrow}
\begin{document}
\title{The Cost of Quantum State Conversion}
\author{Ansh Mishra}
\email{ansh.mishra2025@gmail.com}
\affiliation{Independent researcher, Cumming, Georgia, USA}
\author{Aryan Senthilkumar}
\email{aryan.senthilk@gmail.com}
\affiliation{Independent researcher, Johns Creek, Georgia, USA}
\date{November 14, 2026}
\begin{abstract}
Which losses in quantum-state preparation can better control remove, and which are imposed by the available resources? We determine the least entropy production for converting known full-rank states using fresh thermal auxiliaries of fixed size and bounded energy range, with arbitrary driven joint control. For every reachable target, the asymptotic infimum is the decrease of a computable function of the state's spectrum. Explicit finite protocols approach this limit, including at saturated resource constraints. The result therefore supplies an attainable benchmark for state preparation, rather than only a lower bound. Reusable memory changes both reachability and the cost without enlarging any fresh auxiliary. A uniform memory gives an exact rescaling of the resource parameter; all memory states preparable from it under the same constraint have the same reachable targets and asymptotic costs. Externally supplied biased memories can do better, revealing a preparation resource distinct from dimension. With sufficient fixed memory, protocols reach exact product endpoints while dissipation decreases inversely with operation count. These results connect algorithmic cooling, quantum resource theories, and constructive control by separating unavoidable loss from the overhead of a chosen implementation.
\end{abstract}
\maketitle
\raggedbottom

\section{Introduction}
Preparing a specified quantum state raises two different questions: whether the available resources can reach the target, and how much irreversible loss the conversion must incur. A cooling protocol can achieve its desired polarization without settling the second question. Likewise, improving a control sequence does not show how much further improvement is possible. An informative benchmark must identify the loss imposed by the resource constraint, distinguish it from the overhead of a chosen sequence, and show whether the limiting cost is attainable.

We answer these questions for a controlled system interacting sequentially with fresh auxiliaries of fixed finite dimension and bounded eigenvalue ratio. For thermal auxiliaries, the ratio bounds their entire energy range. The joint unitaries may be driven, and each auxiliary state may be chosen separately. For every reachable pair of known full-rank states, we obtain the exact entropy-production infimum as a difference of endpoint potentials and construct finite protocols approaching it. The lower bound includes arbitrary coherent joint control. Thus a favorable protocol is compared with an attainable limit of the operation class, not merely with another protocol or a potentially loose inequality.

The distinction is useful to several communities for different reasons. For algorithmic cooling, it supplements a reachable polarization with the least irreversible cost of a prescribed conversion, including nonthermal endpoints. For quantum control, it separates an irreducible resource penalty from excess loss that a better implementation may reduce. For resource theory, it identifies how a retained memory changes both the accessible targets and their costs. Section~\ref{sec:implications} makes these uses concrete through a protocol-overhead diagnostic and a known-state reset example. They are benchmarks within an explicit control model, not predictions of device-wide energy savings.

The spectral ingredients have important antecedents. Heat-bath algorithmic cooling establishes geometric limiting states and exponential polarization enhancement \cite{raeisi,tarantoMemory}. Raeisi and Mosca also describe arbitrary initial spectra through expanding and merging consecutive blocks \cite{raeisi}. Bounds based on a machine's largest energy gap extend this picture to other control paradigms \cite{clivaz}, and thermodynamic analyses compare the work and dissipation of particular cooling protocols \cite{lin,tarantoCooling}. Relative-entropy potentials already bound constrained entropy production \cite{kwconstraints}, while passivity deformation supplies tight finite-system inequalities \cite{uzdin}. Discrete quasistatic processes, work extraction, and recovery-based bounds provide complementary approaches \cite{anders,skrzypczyk,alhambra}. The additional assertion here is a matching construction for every reachable full-rank target under the same per-auxiliary cap, including saturated gaps. We use the cooling-limit structure rather than presenting its envelope as an unrelated new object.

Retained memory changes the limiting cost itself. A uniform memory rescales the endpoint parameter without increasing the spectral range of a fresh auxiliary. More generally, every memory state preparable from a uniform register under the same resources has identical product-return reachability and asymptotic costs. An externally supplied biased state can outperform this class only by providing an additional preparation resource; an exact counterexample shows why dimension alone is insufficient. With enough fixed memory, explicit protocols attain exact product return while their dissipation vanishes as the collision count grows. Memory assistance, catalysis, and product return already have substantial antecedents \cite{henao,thermalRecall,liu,wilming,mueller,catalyticthermal}. Our contribution is the attainable cost law and its precise dependence on those memory resources.

Sections~\ref{sec:model} and \ref{sec:law} define the model and establish the law. Sections~\ref{sec:memory}--\ref{sec:coherence} treat memory preparation, finite implementations, and the role of coherent control. The discussion separates their uses for known-state preparation from the additional assumptions needed for hardware-specific control and error-correction tasks. The complete proofs and computational checks are included in the appendices.

\section{Operations and the cost being optimized}\label{sec:model}
Let $S$ have finite dimension $d\ge2$, known full-rank input $\rho$, and known full-rank target $\omega$. Write their eigenvalues in decreasing order as $p$ and $q$. At each collision a fresh, independent auxiliary $A$ has fixed dimension $m\ge2$ and a full-rank state $\tau$ satisfying
\begin{equation}
 \kappa(\tau):=\frac{\lambda_{\max}(\tau)}{\lambda_{\min}(\tau)}\le R,\qquad R>1.\label{eq:cap}
\end{equation}
An arbitrary joint unitary produces $s'=U(\rho\otimes\tau)U^\dagger$ and $\rho'=\tr_A s'$. System-only rotations are allowed, so the endpoint optimization is spectral. Auxiliaries are not reused or jointly batched. There is no external randomization register, feedback controller, or inaccessible reference correlated with the input.

In the thermal interpretation, $\tau=e^{-\beta H_A}/Z_A$ at fixed $\beta>0$, and the cap is the bound $\beta(E_{\max}-E_{\min})\le\log R$ on the entire auxiliary energy range. $H_A$ is fixed during its collision, but it may differ between fresh auxiliaries. The unitaries need not conserve the bare energy.

The collision entropy production is \cite{reeb}
\begin{align}
 \sigma&=D(s'\Vert\rho'\otimes\tau)\nonumber\\
 &=I(S:A)_{s'}+D(\tau'\Vert\tau)\ge0,\label{eq:sigma}\\
 &=\beta Q_A+S(\rho')-S(\rho).\label{eq:heat}
\end{align}
We use natural logarithms and $k_B=1$. The information-theoretic definition does not require thermality. With interactions off at the endpoints and a fixed system Hamiltonian,
\begin{align}
 \Sigma(\Pi)&:=\sum_n\sigma_n=\beta(W-\Delta F_\beta),\nonumber\\
 F_\beta(\rho)&=\tr(H_S\rho)-\beta^{-1}S(\rho).\label{eq:work}
\end{align}
$W$ is the work delivered during the modeled collisions. This accounting excludes auxiliary-preparation infrastructure, gate synthesis, elapsed time, noise suppression, and apparatus manufacture.

We optimize in endpoint closure:
\begin{equation}
 \cost_{R,m}(p\to q)=
 \lim_{\eps\downarrow0}\ \inf_{\Pi:\,\frac12\|p_\Pi-q\|_1\le\eps}\Sigma(\Pi).
 \label{eq:cost}
\end{equation}
Here $p_\Pi$ is the sorted output spectrum; the infimum of the empty set is $+\infty$. Protocol lengths may grow, while $d,m,R$ remain fixed. Output rotations give convergence to the specified target eigenbasis. This is an asymptotic infimum, not the optimum at a prescribed collision count. In particular, a nontrivial change of spectrum cannot have exactly zero entropy production in a finite protocol: $\sigma=0$ implies $s'=\rho'\otimes\tau$, whose spectrum and all power traces agree with those of $\rho\otimes\tau$, forcing $\rho'$ and $\rho$ to have the same spectrum.

\section{An exact dissipation law}\label{sec:law}
For decreasing normalized vectors, $p\prec r$ means $\sum_{i\le k}p_i\le\sum_{i\le k}r_i$ for each $k<d$. Call $r$ \emph{$R$-separated} if $r_i\ge Rr_{i+1}>0$. Let $\B_R(p)$ be the least $R$-separated majorization upper bound of $p$.

This envelope can be calculated directly. Define
\begin{equation}
 \gamma_{d,R,i}=\frac{R^{d-i}}{\sum_{j=1}^d R^{d-j}},\quad
 \Gamma_k=\sum_{i\le k}\gamma_{d,R,i},\quad P_k=\sum_{i\le k}p_i.\label{eq:gamma}
\end{equation}
Let $L_p$ be the least concave majorant of $(\Gamma_k,P_k)$, including $(0,0)$ and $(1,1)$. Then
\begin{equation}
 \B_R(p)_i=L_p(\Gamma_i)-L_p(\Gamma_{i-1}).\label{eq:envelope}
\end{equation}
Equivalently, merge adjacent blocks whenever their mass-to-geometric-weight slopes increase, merging ties. On a final block $B$,
\begin{equation}
 \B_R(p)_i=\frac{P_B}{G_B}\gamma_{d,R,i},\quad
 P_B=\sum_{i\in B}p_i,\quad G_B=\sum_{i\in B}\gamma_{d,R,i}.\label{eq:blocks}
\end{equation}
This is a standard isotonic construction \cite{lim}. It takes $O(d)$ arithmetic operations after sorting, with unit-cost arithmetic understood. Its geometric blocks express the asymptotic cooling structure discussed in Ref.~\cite{raeisi}.

The potential that determines the cost is
\begin{align}
 \Phi_R(p)&=D\bigl(\B_R(p)\Vert\gamma_{d,R}\bigr)\nonumber\\
 &=\sum_B P_B\log(P_B/G_B).\label{eq:phi}
\end{align}
The dimension is implicit in the argument of $\Phi_R$.

\begin{theorem}[Exact asymptotic dissipation]\label{thm:law}
For full-rank $p,q$, $R>1$, and every fixed $m\ge2$,
\begin{equation}
 \boxed{\cost_{R,m}(p\to q)=
 \begin{cases}
  \Phi_R(p)-\Phi_R(q),&q\prec\B_R(p),\\
  +\infty,&q\not\prec\B_R(p).
 \end{cases}}\label{eq:law}
\end{equation}
The finite value is approachable by admissible finite protocols. No optimum for a fixed number of collisions is asserted.
\end{theorem}

The theorem covers arbitrary coherent joint unitaries and arbitrary intermediate paths. It also yields the reachability and zero-dissipation criteria:
\begin{align}
 p\cl_R q&\ \Longleftrightarrow\ q\prec\B_R(p),\label{eq:reach}\\
 \cost_{R,m}(p\to q)=0&\ \Longleftrightarrow\ \B_R(p)=\B_R(q).\label{eq:zero}
\end{align}
Both are independent of the fixed fresh-auxiliary dimension. Implementation costs and convergence rates need not be.

\subsection{The lower bound}
For an $R$-separated spectrum $r$, let $\C_r=\{\eta:\lambda(\eta)\prec r\}$. This set is invariant under every admissible collision and its reset reverse
\begin{equation}
 \mathcal R(\eta)=\tr_A[U^\dagger(\eta\otimes\tau)U].\label{eq:reverse}
\end{equation}
Indeed, separation orders the rows of $r\otimes\lambda(\tau)$. Its $km$ largest eigenvalues sum to the first $k$ entries of $r$, and the Ky Fan variational principle bounds the corresponding output-system sum. This is the cooling-ceiling argument \cite{clivaz}, applied to every separated reference. The reset reverse is not identified with a channel inverse or, in general, a Petz map \cite{aw}.

For any reverse-invariant set $\C$, put $E_\C(\rho)=\inf_{\eta\in\C}D(\rho\Vert\eta)$. The chain identity and data processing give
\begin{align}
 \sigma+D(\rho'\Vert\eta)
 &=D\bigl(\rho\otimes\tau\Vert U^\dagger(\eta\otimes\tau)U\bigr)\nonumber\\
 &\ge D(\rho\Vert\mathcal R(\eta))\ge E_\C(\rho).
 \label{eq:reversebound}
\end{align}
Taking the infimum and summing proves
\begin{equation}
 \Sigma(\Pi)\ge E_\C(\rho_0)-E_\C(\rho_\Pi).\label{eq:telescoping}
\end{equation}
For $\C=\C_{\gamma_{d,R}}$, the projection formula in Appendix~\ref{app:envelope} yields
\begin{equation}
 E_{\C_{\gamma_{d,R}}}(p)=\Phi_R(p).\label{eq:projectionphi}
\end{equation}
The minimizing reference state is generally not the envelope itself. Both divergences nevertheless reduce to Eq.~\eqref{eq:phi}. Continuity then gives the lower bound in Eq.~\eqref{eq:law}. The relative-entropy form has clear antecedents \cite{kwconstraints,alhambra}; attainability for every reachable target in this collision model is the additional assertion.

\subsection{Attainability, including saturated gaps}
Put $a=\B_R(p)$ and $b=\B_R(q)$. Reachability implies $b\prec a$. We first show in Appendix~\ref{app:zero} that any two equal-envelope states are connected in closure at vanishing cost. It therefore remains to connect $a$ to $b$.

The straight path $x(t)=(1-t)a+tb$ stays $R$-separated. Define its downward cumulative flows
\begin{equation}
 F_i=\sum_{j\le i}(a_j-b_j)\ge0,\qquad 1\le i<d.\label{eq:flows}
\end{equation}
On each of $n$ slices, move $M_i=F_i/n$ from level $i$ to $i+1$. Sampled endpoints are exact; intermediate states are only $O(n^{-1})$ from the separated path and need not themselves be separated.

\begin{figure}[tbp]
\includegraphics[width=\columnwidth]{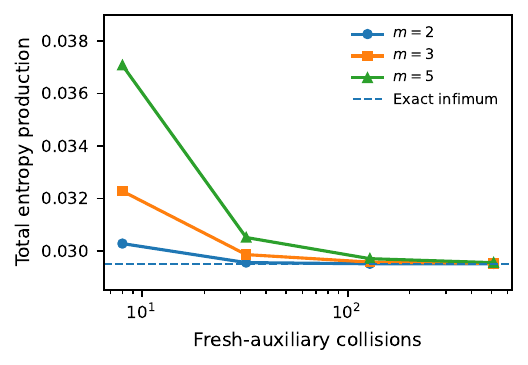}
\caption{Approaching the exact cost at saturated gaps. The endpoints are $a=(0.8,0.15,0.05)$ and $b=\gamma_{3,3}=(9,3,1)/13$. Each point uses exact rational controls and endpoints; costs are high-precision evaluations. The horizontal dashed line is the common infimum $0.0295083998304\ldots$. Larger auxiliary dimension changes this construction's finite overhead, not the limiting law. Plotting code was AI-assisted.}\label{fig:middle}
\end{figure}

For an actual incoming pair $(A,B)$ and a downward transfer $M>0$, take
\begin{align}
 t_*&=\frac{A-(m-1)M}{B+M},\qquad t=\min\{R,t_*\},\nonumber\\
 \tau_0&=\frac{t}{t+m-1},\qquad \tau_j=\frac1{t+m-1}\quad(j\ge1).
 \label{eq:downcontrol}
\end{align}
Rotate only $|i,1\rangle$ and $|i+1,0\rangle$, with
\begin{equation}
 \sin^2\theta=\frac{M(t+m-1)}{A-tB}.\label{eq:angle}
\end{equation}
For a sufficiently fine mesh, $t_*>1$ and the expression lies in $(0,1]$. The transfer is exact, the auxiliary satisfies its cap, and
\begin{equation}
 \sigma=\Delta S-M\log t.\label{eq:downcost}
\end{equation}
Even at a saturated gap, $R-t=O(n^{-1})$ whenever retuning is needed. Thus each correction $M\log(R/t)$ is nonnegative and $O(n^{-2})$. Summing gives
\begin{align}
 \Sigma_n(a\to b)
 &=H(b)-H(a)-\log R\sum_i F_i\nonumber\\
 &\quad+\sum_{\text{steps}}M\log(R/t)\nonumber\\
 &=D(a\Vert\gamma_{d,R})-D(b\Vert\gamma_{d,R})+O(n^{-1}).
 \label{eq:middlecost}
\end{align}
Appendix~\ref{app:attainment} supplies the uniform estimates and the order of limits joining the outer closure stages. Figure~\ref{fig:middle} tests the construction at a boundary.

For separated $a\succ b$,
\begin{equation}
 D(a\Vert\gamma)-D(b\Vert\gamma)
 =D(a\Vert b)+\sum_i(a_i-b_i)\log\frac{b_i}{\gamma_i}
 \ge D(a\Vert b).\label{eq:strict}
\end{equation}
Summation by parts proves the inequality because $b_i/\gamma_i$ is nonincreasing. Distinct reachable envelopes therefore have strictly positive optimal cost, recovering Eq.~\eqref{eq:zero}.

For example, at $R=3$ the qubit conversion $(.85,.15)\to(.80,.20)$ has exact infimum $0.02276272\ldots$, compared with the target-specific bound $0.00837862\ldots$. For $(.8,.12,.08)\to(.7,.18,.12)$, the infimum is $0.02936886\ldots$, whereas that bound gives $0.02573209\ldots$. Both values follow directly from Eq.~\eqref{eq:phi}; a fixed finite protocol need not attain them.

\section{Returned memory and its preparation}\label{sec:memory}
Retain a memory $C$ of fixed dimension $K$ in a full-rank state $c$. Treat $SC$ as the controlled system, allowing arbitrary $SCA$ unitaries while each fresh auxiliary keeps dimension $m$ and cap $R$. The joint output must approach $q\otimes c$, not merely return the memory marginal. Denote this closure cost by $\cost^c_{R,m}(p\to q)$. Theorem~\ref{thm:law} applies to the sorted spectra of $p\otimes c$ and $q\otimes c$.

The cost is computed on $SC$. At product endpoints,
\begin{equation}
 \Sigma=\beta Q_{\rm tot}+S(q)-S(p).\label{eq:memoryheat}
\end{equation}
Temporary memory entropy and correlations are included. Retaining all spent auxiliaries mathematically as $E$ gives
\begin{equation}
 \Sigma=I(SC:E)_{\rm final}
   +D(\rho_E^{\rm final}\Vert\tau_1\otimes\cdots\otimes\tau_N).
 \label{eq:global}
\end{equation}
Exact product return below concerns $S$ and $C$. At finite $N$ the memory may remain correlated with spent auxiliaries; those correlations are counted by Eq.~\eqref{eq:global}, and $I(C:E)\le\Sigma$.

\subsection{Uniform memory}
Let $u_K=\id_K/K$. The geometric reference factorizes as $\gamma_{dK,R}=\gamma_{d,R^K}\otimes\gamma_{K,R}$, in decreasing lexicographic order. The envelope has the matching identity
\begin{equation}
 \B_R(p\otimes u_K)=\B_{R^K}(p)\otimes\gamma_{K,R}.\label{eq:uniform}
\end{equation}
Appendix~\ref{app:memory} proves this by grouping the repeated input eigenvalues in the concave-majorant construction. Consequently,
\begin{equation}
 \Phi_R(p\otimes u_K)=\Phi_{R^K}(p).\label{eq:uniformphi}
\end{equation}

\begin{theorem}[Uniform-memory cost]\label{thm:uniform}
A target $q\otimes u_K$ is reachable in closure from $p\otimes u_K$ exactly when $q\prec\B_{R^K}(p)$. For such endpoints,
\begin{equation}
 \boxed{\cost^{u_K}_{R,m}(p\to q)=\Phi_{R^K}(p)-\Phi_{R^K}(q).}\label{eq:uniformcost}
\end{equation}
\end{theorem}

\begin{figure}[tbp]
\includegraphics[width=\columnwidth]{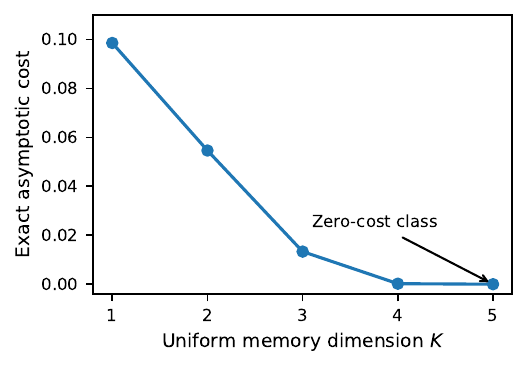}
\caption{Uniform memory reduces the exact asymptotic cost for $(.99,.01)\to(.95,.05)$ at fresh-auxiliary cap $R=3$. The values for $K=1,\ldots,5$ are $0.0985692$, $0.0546247$, $0.0133245$, $0.000213050$, and zero. Zero is a vanishing-cost infimum, not a nontrivial zero-cost finite protocol. Plotting code was AI-assisted.}\label{fig:memorycost}
\end{figure}

The substitution changes the endpoint test and the positive optimum without changing the cap of any physical fresh auxiliary. It does not equate channels or finite-collision costs. $K$ denotes dimension: for $b$ memory qubits, $K=2^b$. Exponential cooling enhancement is established \cite{raeisi,tarantoMemory}; Eq.~\eqref{eq:uniformcost} concerns the full asymptotic dissipation cost with returned product memory. Figure~\ref{fig:memorycost} shows an example.

\subsection{All preparable memories are equivalent}
The appropriate comparison depends on how the retained state is supplied. By Eq.~\eqref{eq:reach}, the states preparable in closure from $u_K$ with cap $R$ are exactly
\begin{equation}
 c\prec\gamma_{K,R}.\label{eq:preparable}
\end{equation}
Their preparation and its reverse have vanishing entropy-production infimum. This does not mean zero work: reversible preparation can change the memory's free energy.

\begin{theorem}[Equivalence of preparable memories]\label{thm:preparable}
If $c\prec\gamma_{K,R}$, then for every full-rank $p$,
\begin{equation}
 \B_R(p\otimes c)=\B_R(p\otimes u_K)
   =\B_{R^K}(p)\otimes\gamma_{K,R}.\label{eq:memoryequal}
\end{equation}
Thus $c$ and $u_K$ have identical product-return reachability and closure costs:
\begin{equation}
 \cost^c_{R,m}(p\to q)=\cost^{u_K}_{R,m}(p\to q),\label{eq:memorycostequal}
\end{equation}
including $+\infty$ for unreachable endpoints.
\end{theorem}
\begin{proof}
Put $b=\B_{R^K}(p)$. Since $u_K\prec c\prec\gamma_{K,R}$ and $p\prec b$,
\begin{equation}
 p\otimes u_K\prec p\otimes c\prec b\otimes\gamma_{K,R}.
\end{equation}
The last vector is $R$-separated and, by Eq.~\eqref{eq:uniform}, equals $\B_R(p\otimes u_K)$. Monotonicity and minimality of the envelope squeeze the middle envelope to this same vector. For a target, $q\prec b$ implies $q\otimes c\prec b\otimes\gamma_{K,R}$. Conversely, the largest $kK$ entries of $q\otimes c$ sum to at least $\sum_{i\le k}q_i$, by selecting all memory entries from those $k$ system levels. The corresponding prefix of $b\otimes\gamma_{K,R}$ is $\sum_{i\le k}b_i$, so product reachability implies $q\prec b$. Equation~\eqref{eq:memorycostequal} now follows from Theorem~\ref{thm:law} and the factorized reference.
\end{proof}

This equality is stronger than uniform-memory optimality. It also avoids having to compose approximate memory-preparation protocols: the equivalence follows directly from an algebraic sandwich. It is an endpoint-closure statement; finite operation counts and exact-return properties at boundary points may differ.

For an arbitrary full-rank memory, let
\begin{equation}
 T(c;R)=\min\{T\ge R:c\prec\gamma_{K,T}\}.\label{eq:memorybudget}
\end{equation}
For $K=1$ take $T=R$. The minimum exists for positive $c$. The same tensor argument gives the necessary condition
\begin{equation}
 p\otimes c\cl_R q\otimes c\ \Longrightarrow\ q\prec\B_{T^K}(p).
 \label{eq:arbitrarymemory}
\end{equation}
Enlarging the fresh-auxiliary cap from $R$ to $T$ also gives, for an $R$-reachable pair,
\begin{equation}
 \cost^c_{R,m}(p\to q)\ge\Phi_{T^K}(p)-\Phi_{T^K}(q).\label{eq:arbitrarycost}
\end{equation}
These are bounds, not equivalences for general externally supplied memories.

\subsection{Why the preparation restriction is necessary}
Take the exact rational data
\begin{align}
 R&=3,\qquad c=(4/5,1/5),\nonumber\\
 p&=(.868,.1206,.0084,.0030),\nonumber\\
 q&=(.868,.1205,.0085,.0030).\label{eq:biasedexample}
\end{align}
The joint envelopes coincide:
\begin{equation}
 \B_3(p\otimes c)=\B_3(q\otimes c)
  =\left(\frac{434}{625}\right)\oplus\frac{191}{625}\gamma_{7,3}.
 \label{eq:biasedenvelope}
\end{equation}
All proper internal prefixes of the seven-entry block are strict for both endpoints; the smallest unnormalized slack is $1181/5465000$. The exact-return construction below therefore applies.

Uniform two-level memory does not give zero cost:
\begin{align}
 \B_9(p)&=(.88974,.09886,.01026,.00114),\nonumber\\
 \B_9(q)&=(.88965,.09885,.01035,.00115).\label{eq:biasedcoarse}
\end{align}
The forward uniform-memory cost is $6.38035161148\times10^{-6}$, and the reverse direction is unreachable. Uniform three-level memory suffices. However, $c\not\prec\gamma_{2,3}=(3/4,1/4)$; it has $T=4$, not $3$. Neither $p\otimes u_2\to p\otimes c$ nor $q\otimes u_2\to q\otimes c$ is reachable at cap $3$. The biased memory is a borrowed preparation resource, not a contradiction of Theorem~\ref{thm:preparable} or an equal-purity comparison.

\section{Exact return and finite implementations}\label{sec:finite}
The preceding cost laws concern endpoint closure. A stronger construction is possible when there is slack.
\begin{proposition}[Exact return at fixed memory size]\label{prop:exact}
For any two specified full-rank states, $R>1$, and any fixed fresh-auxiliary dimension $m\ge2$, a sufficiently large but fixed uniform memory admits finite protocols with
\begin{equation}
 \rho\otimes u_K\longrightarrow\omega\otimes u_K,
 \qquad\Sigma_N=O(N^{-1}).\label{eq:exactreturn}
\end{equation}
Both endpoints are exact products in every completed protocol. The same $K$ works in the reverse direction.
\end{proposition}
If $\eps$ is the smallest eigenvalue of either endpoint, the strict inequality $R^{-K}<\eps$ is sufficient. It makes both repeated spectra strictly majorized by $\gamma_{dK,R}$. Appendix~\ref{app:zero} constructs a path with a positive cap margin and implements its sampled increments exactly; no unrestricted final swap is used. The return error is zero, while $K$ remains fixed as $N$ grows. The controls are tailored to known states, not a channel acting correctly on all unknown inputs.

The minimum uniform-memory dimension for a vanishing-cost \emph{closure} is
\begin{equation}
 K_{\min}(p,q;R)=\min\{K\ge1:\B_{R^K}(p)=\B_{R^K}(q)\}.\label{eq:kmin}
\end{equation}
Theorem~\ref{thm:preparable} makes this minimum valid over all memories satisfying Eq.~\eqref{eq:preparable}, not over arbitrary borrowed states. Monotonicity in $K$ follows from $\B_T(\B_R(p))=\B_T(p)$ for $T\ge R$.

For distinct qubit spectra with largest eigenvalues $x,y\ge1/2$,
\begin{equation}
 K_{\min}=\left\lceil\frac{\log[x_*/(1-x_*)]}{\log R}\right\rceil,
 \qquad x_*=\max(x,y).\label{eq:qubitthreshold}
\end{equation}
Equality at the threshold refers to closure; strict inequality in the resource ratio supplies exact return. More generally, for all known $d$-level states with eigenvalues at least $\eps$, the sharp common closure budget is the smallest $K$ such that
\begin{equation}
 \gamma_{d,R^K,1}\ge1-(d-1)\eps.\label{eq:domain}
\end{equation}
The extremal spectrum $(1-(d-1)\eps,\eps,\ldots,\eps)$ witnesses necessity. Sufficiency follows by majorization; strict inequality guarantees exact return for every pair. These budgets do not bound runtime or control complexity.

\subsection{A qubit separation with an improved schedule}
At $R=3$, consider $p=(.85,.15)$ and $q=(.80,.20)$. Without memory, $q\to p$ is unreachable. The forward optimum is
\begin{align}
 \cost_{3,m}(p\to q)
 &=d_{\rm B}(.85\Vert.75)-d_{\rm B}(.80\Vert.75)\nonumber\\
 &=0.0227627212988\ldots .\label{eq:floor}
\end{align}
This is the qubit specialization of Eq.~\eqref{eq:law}; its heat-bound ingredients are prior results \cite{uzdin,clivaz}. A two-level auxiliary and one partial rotation attain this particular positive optimum.

With one uniform memory qubit, use the joint-population path
\begin{align}
 P&=(.425,.425,.075,.075),&A&=(.60,.25,.11,.04),\nonumber\\
 B&=(.55,.25,.14,.06),&Q&=(.40,.40,.10,.10).
 \label{eq:path}
\end{align}
The order is $(00,01,10,11)$. The first and last segments transfer within system rows; the middle segment interleaves adjacent transfers. Every pair move $(a,b)\mapsto(a+M,b-M)$ uses
\begin{align}
 \tau&=\left(\frac{a+M}{a+b},\frac{b-M}{a+b}\right),\nonumber\\
 \sigma&=a\log\frac a{a+M}+b\log\frac b{b-M},\label{eq:paircost}
\end{align}
and the transposition $|j,0\rangle\leftrightarrow|i,1\rangle$. These local controls are related to collision-based cooling constructions \cite{tarantoCooling,mishraSenthilkumarCooling}; here they implement product-memory return.

For equal subdivision of the seven transfer strokes, $N=7n$. Both directions satisfy the cap for $n\ge5$, and the exact cost at $N=63$ first falls below Eq.~\eqref{eq:floor} among the admissible counts tested, whereas $N=56$ does not. The conservative analytical bound $8589/(400N)$ proves vanishing cost but certifies a crossing only at $N=945$ within this family. The earlier crossing follows from directed evaluation of Eq.~\eqref{eq:paircost}, not from that coarse bound.

We improve the finite schedule without changing the macroscopic path. Five segments are sampled at uniform local-metric arclength: two pair strokes, the interleaved three-edge segment, and two final pair strokes. Collisions are allocated between segments in proportion to their lengths. The resulting leading coefficient is approximately $1.14928$, compared with $1.387$ for equal subdivision. This is a restricted scheduling improvement, not a global path optimum. Appendix~\ref{app:finite} describes the integer allocation and rational sampling.

\begin{table}[tbp]
\caption{Forward exact-return qubit protocols on the same macroscopic path. All controls and product endpoints are rational and exactly checked. The improved $N=56$ and $70$ comparisons with the memoryless floor have directed interval certificates. Other displayed costs are 70-digit numerical evaluations rounded here.}\label{tab:protocol}
\begin{ruledtabular}
\begin{tabular}{rcc}
$N$ & Equal subdivision & Reallocated collisions\\
\hline
56 & 0.024756063453 & 0.020424801302\\
70 & 0.019809236626 & 0.016360931939\\
280 & 0.004955674219 & 0.004100221229\\
1120 & 0.001239133497 & 0.001025874354\\
2240 & 0.000619584754 & 0.000513003878
\end{tabular}
\end{ruledtabular}
\end{table}

\begin{figure}[tbp]
\includegraphics[width=\columnwidth]{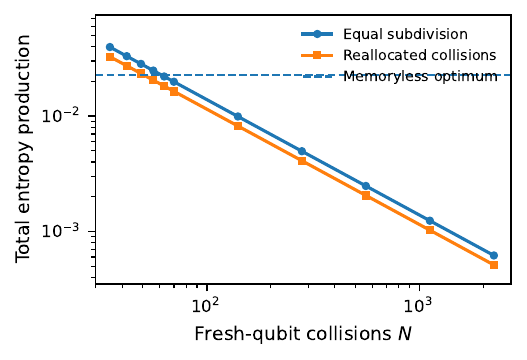}
\caption{Finite overhead on a zero-infimum memory-assisted conversion. Both schedules return the uniform memory exactly in product with the target. The horizontal dashed line is the positive optimum without retained memory, Eq.~\eqref{eq:floor}. Reallocation beats it at $N=56$; equal subdivision requires $N=63$ in the tested family. Points are evaluations of the saved rational controls. Plotting code was AI-assisted.}\label{fig:finite}
\end{figure}

\section{What coherent control changes}\label{sec:coherence}
Theorem~\ref{thm:law} establishes optimality against arbitrary coherent joint control. Its infimum is approachable while the reduced controlled-system states remain diagonal between collisions, but this does not eliminate all joint coherence. Partial rotations in Eq.~\eqref{eq:angle} can be essential at fixed auxiliary dimension.

Restrict each collision to a permutation of a product eigenbasis, with no externally randomized mixtures. Let a qubit have largest eigenvalue $x>R/(1+R)$. Every nontrivial spectral change then satisfies
\begin{equation}
 x-x'\ge\frac{x-R(1-x)}{1+(m-1)R}.\label{eq:permgap}
\end{equation}
All upper-row joint populations exceed all lower-row ones. A mixed row selection replaces at least one upper entry by a lower entry, losing at least $\tau_{\min}[x-R(1-x)]$. Normalization and the cap imply $\tau_{\min}\ge[1+(m-1)R]^{-1}$. The same argument applies after exchanging output rows; Appendix~\ref{app:perm} treats sequential use.

For $x=.85$ and $R=3$, the highest nontrivial landing point is $.75$ for $m=2$ and $111/140=.792857\ldots$ for $m=3$. Thus neither permutation-only model can reach $.80$, even after arbitrarily many steps. A coherent two-state rotation with $\tau=(.75,.25)$ reaches it in one step and attains Eq.~\eqref{eq:floor}. With retained uniform memory, the zero-cost construction instead uses transpositions on a larger controlled system.

This is a separation from deterministic product-eigenbasis permutations, not from all classical stochastic dynamics. Quantum advantages for fixed finite baths have been studied in other operation classes \cite{scharlau}; Eq.~\eqref{eq:permgap} specifies the restricted comparison relevant here.

\section{Implications for control and reusable quantum resources}\label{sec:implications}
\subsection{Separating a resource limit from control overhead}
For an admissible finite protocol $\Pi$ reaching a specified target exactly, define its overhead above the law by
\[
 \mathcal E(\Pi;p,q)
 :=\Sigma(\Pi)-[\Phi_R(p)-\Phi_R(q)]\ge0.
\]
This is a direct consequence of Eq.~\eqref{eq:telescoping}, not a separate optimality assumption. The potential difference is fixed by the endpoints and resource constraint; $\mathcal E$ depends on the implementation. A better schedule can reduce this overhead, as Table~\ref{tab:protocol} illustrates, but cannot remove a positive resource-imposed cost while keeping the same operation class. For an approximate target, the rigorous comparison uses the actual output spectrum; Appendix~\ref{app:envelope} bounds the correction associated with target error.

This gives a concrete use for the law in control design. Specify the known input and target, auxiliary size, and spectral cap; evaluate reachability and the potential difference; then compare the cost of any admissible candidate sequence with that value. A large overhead identifies room left by the thermodynamic benchmark, not a guarantee that a particular pulse optimizer can realize the improvement. Conversely, approaching the benchmark shows that substantial further reduction requires changing the resource model rather than merely retiming the same task.

A hardware model with fewer allowed gates remains subject to this lower bound \emph{if} every operation and its cost are contained in the present collision model. Attainability need not survive those extra restrictions. Additional dissipative channels, correlated resets, or control-dependent environments must be accounted for separately before importing the bound. Our explicit operations provide a constructive reference, not a compiled pulse program.

\subsection{From cooling limits to a known-state reset budget}
The distinction can be seen without a full protocol search. Consider preparing the qubit state $q_\epsilon=(1-\epsilon,\epsilon)$ from $u_2$, with $0<\epsilon<1/2$. Here $\epsilon$ is the unwanted-state population, not the numerical accuracy of the optimization. Equations~\eqref{eq:reach} and \eqref{eq:uniformcost} give
\[
 u_2\otimes u_K\cl_R q_\epsilon\otimes u_K
 \quad\Longleftrightarrow\quad
 \epsilon\ge\frac{1}{1+R^K}.
\]
Whenever this condition holds, the conversion has zero entropy-production infimum: both coarse endpoints have the same geometric envelope. A strict inequality supplies exact product-return protocols through the strict-slack construction in Appendix~\ref{app:zero}. At equality, the statement remains one of closure.

For example, at $R=3$ a maximally mixed qubit cannot be prepared with target population $0.99$ without retained memory. A uniform five-level memory suffices, whereas four levels do not: $3^4=81<99<243=3^5$. The five-level register stays fixed as the operation count increases, and it is returned in product with the target. This is a specialization of Eq.~\eqref{eq:qubitthreshold}, not a new cooling ceiling or a finite-time performance prediction. $K$ counts memory levels, not memory qubits.

The example concerns a specified, initially uncorrelated input. It is not a reset channel for every unknown input, nor a thermodynamic analysis of an error-correction loop. Extending it to such a loop would require explicit treatment of any correlations with a data register, measurement and feedback records, reused resources, and timing. Exact pure-state reset is also outside the full-rank theorem. These distinctions identify what a computing application would need beyond the present benchmark.

\subsection{Why memory preparation belongs in resource accounting}
Theorem~\ref{thm:preparable} makes uniform and preparable biased memories equivalent in closure, while Eq.~\eqref{eq:biasedexample} identifies a borrowed state outside that class. Thus dimension alone does not specify a memory's thermodynamic value. Equivalence of endpoint costs also does not imply equal pulse complexity or equal cost at a fixed collision count. This differs from correlating-catalyst entropy criteria \cite{wilming,mueller} and energy-conserving catalytic thermal operations \cite{catalyticthermal}. \emph{Thermal Recall} already restores a thermal memory in product form by final thermalization \cite{thermalRecall}; no omitted cleanup is attributed to that framework.

Retained memory is not the only way to enlarge accessible spectral ratios. A batch of $b$ independent $m$-level auxiliaries has dimension $m^b$ and condition number at most $R^b$. For $b=K$ this is the parameter appearing in Eq.~\eqref{eq:uniformcost}, but a fresh batch and a retained $K$-level register have different resource budgets. Product batches are also a restricted subset of all larger auxiliaries with that cap. The ratio identity alone does not establish channel or finite-cost equivalence.

\subsection{What is exact, and what remains device dependent}
The law concerns finite-size auxiliaries with an unrestricted number of collisions. It is not a universal finite-time correction to quantum computation. The matching construction distinguishes it from a lower bound whose attainability is unknown; the explicit finite examples show that implementation overhead remains even when the asymptotic cost vanishes. Optimizing that overhead, noise tolerance, gate locality, and elapsed time remains separate work.

The theoretical connection is therefore specific: cooling-limit structure and constrained thermodynamic potentials become an attainable benchmark for known-state preparation and for the use of returned memory. Geometric ceilings and block merging \cite{raeisi,clivaz}, potential-difference bounds \cite{kwconstraints}, and the qubit passivity inequality \cite{uzdin} are established ingredients. The present result identifies the exact cost and an approaching construction for all reachable full-rank endpoints in the stated operation class. It does not establish an experimental energy saving, an advantage in sensing, or a device-level error-correction improvement.

\section{Conclusion}
The least entropy production in bounded-auxiliary state conversion is a computable difference of spectral potentials. A lower bound valid against arbitrary coherent joint control and a matching construction establish the asymptotic law for every reachable full-rank target, including saturated resource boundaries. This separates the cost imposed by available resources from the overhead of a chosen sequence. Returned memory changes the limiting benchmark, and its initial preparation matters independently of dimension. Together, the results give quantum cooling, resource theory, and control a common way to assess specified state transformations before imposing the timing, noise, and interaction constraints of a particular implementation.

\section*{Data and code availability}
The accompanying Supplemental Material \cite{supp} contains source code, saved rational controls, directed interval endpoints, deterministic seeds, software versions, and execution logs. It distinguishes exact algebra, interval-certified inequalities, and numerical regression tests. No experimental measurements were collected.

\section*{Computational methods}
Calculations use Python, NumPy, SciPy, mpmath, and exact integer/rational arithmetic. Tests compare independent envelope constructions, direct joint-unitary evolution, reduced updates, and analytical cost identities. Appendix~\ref{app:checks} records the executed scope and rejected coarse schedules. Tests support the implementations; universal claims rest on the proofs.

ChatGPT (GPT-6 Astra Pro) assisted with theoretical development, literature review, code, tests, and plots under model-specific prompts. Research provenance and computational checks are documented in the Supplemental Material.
\section*{Acknowledgments}
ChatGPT also assisted with drafting and revising the text.

\appendix

\section{Envelope and projection identities}\label{app:envelope}
The slopes of a cumulative curve on the abscissae $\Gamma_i$ are $r_i/\gamma_{d,R,i}$. They are nonincreasing exactly when $r_i\ge Rr_{i+1}$. Thus separated majorants of $p$ are precisely concave cumulative curves above its data. The least concave majorant has positive increments, proving Eq.~\eqref{eq:envelope} and
\begin{gather}
 p\prec\B_R(p),\qquad \B_R(\B_R(p))=\B_R(p),\nonumber\\
 p\prec q\Rightarrow\B_R(p)\prec\B_R(q).\label{eq:basicenv}
\end{gather}
Strict slope changes are contact points. On a maximal geometric block, the input and envelope therefore have equal total mass, and the within-block input is majorized by the geometric block after normalization.

For a decreasing full-rank ceiling $r$, trace rearrangement aligns the minimizing reference eigenbasis with the input in $E_{\C_r}(\rho)$. The classical optimization is $\min_{s\prec r}D(p\Vert s)$. Pool adjacent blocks using nonincreasing slopes $\ell_B=P_B/R_B$, where $R_B=\sum_{i\in B}r_i$. Then
\begin{equation}
 s_i^*=p_i/\ell_B\quad(i\in B),\qquad
 E_{\C_r}(p)=\sum_B P_B\log(P_B/R_B).\label{eq:projection}
\end{equation}
Each internal input prefix is at most $\ell_B$ times its reference prefix, giving feasibility. At a block boundary the last left entry of $s^*$ is at least the corresponding entry of $r$, and the first right entry is at most its reference entry, so $s^*$ is globally sorted. Assign multiplier $\ell_B-\ell_{B+1}\ge0$ at each block-boundary prefix, zero at other prefixes, and normalization multiplier equal to the last slope. Stationarity is $-p_i/s_i^*+\ell_B=0$, with complementary slackness at contacts. Strict convexity proves uniqueness.

At $r=\gamma_{d,R}$, the projection uses exactly the block partition in Eq.~\eqref{eq:blocks}. The envelope $b_i=\ell_B\gamma_i$ and the minimizing reference $s_i^*=p_i/\ell_B$ generally differ, but both $D(b\Vert\gamma)$ and $D(p\Vert s^*)$ equal $\sum_B P_B\log\ell_B$. This proves Eq.~\eqref{eq:projectionphi}.

Every state in $\C_r$ has minimum eigenvalue at least $r_d$. Hence the minimizing set is compact and uniformly full rank. For trace distance $\delta=\frac12\|\rho-\omega\|_1\le1-1/d$, entropy continuity \cite{audenaert} gives the conservative bound
\begin{align}
 |E_{\C_r}(\rho)-E_{\C_r}(\omega)|
 &\le h_2(\delta)+\delta\log(d-1)\nonumber\\
 &\quad+2\delta\log(1/r_d).\label{eq:continuity}
\end{align}
It follows by applying the entropy estimate and bounding the cross-entropy difference for each reference, then taking infima. In particular, the finite-protocol lower bound with $r=\gamma_{d,R}$ remains informative for imperfect target preparation. This continuity is also used when finite protocols are composed below.

\section{Zero-cost connections and exact sampled paths}\label{app:zero}
\subsection{A fixed-dimension local control}
For adjacent controlled-system populations $(a,b)>0$ and $m=k+1$, use the cycle
\begin{equation}
 |j,0\rangle\to|i,1\rangle\to\cdots\to|i,k\rangle\to|j,0\rangle.
\end{equation}
The normalized auxiliary
\begin{align}
 z&=\frac{ab+bM(k+1)/2}{a+kb},\nonumber\\
 \tau_0&=z/b,\qquad \tau_y=(z-yM/k)/a\quad(1\le y\le k)
 \label{eq:cyclecontrol}
\end{align}
implements $(a,b)\mapsto(a+M,b-M)$ exactly. Its active joint populations are $z,z-M/k,\ldots,z-M$, so level $j$ loses $M$ and each of the $k$ active level-$i$ configurations gains $M/k$. At $M=0$, all active weights equal $w=ab/(a+kb)$ and the auxiliary ratio is $\max(a/b,b/a)$. On a compact positive set with strict cap slack, sufficiently small signed $M$ preserves positivity and the cap uniformly.

The exact identity
\begin{equation}
 \sigma=D(s'\Vert\rho\otimes\tau)-D(\rho'\Vert\rho)\label{eq:refund}
\end{equation}
follows from unitary invariance of joint entropy. Expanding relative entropy about the resonant state yields
\begin{equation}
 \sigma=\frac{M^2}{2}\left(\frac{k}{a}+\frac{1}{kb}\right)+O(M^3),\label{eq:cyclecost}
\end{equation}
uniformly on that compact set. Indeed, the joint quadratic term is $M^2(1+1/k)/(2w)$ and the system refund is $M^2(1/a+1/b)/2$. This local primitive and its resonance structure connect to Ref.~\cite{mishraSenthilkumarCooling}; no unrestricted optimality of the primitive is assumed.

\subsection{An exact sampling lemma}
\begin{lemma}\label{lem:sampling}
Let $x(t)$, $0\le t\le T$, be a piecewise continuously differentiable positive population path, with bounded derivative and positive minimum coordinate. Write $X_i(t)=\sum_{j\le i}x_j(t)$. Suppose a constant $R_0<R$ satisfies
\begin{equation}
 \max\{x_i/x_{i+1},x_{i+1}/x_i\}\le R_0
 \quad\text{whenever }\dot X_i\ne0.\label{eq:slack}
\end{equation}
Then exact-endpoint protocols with any fixed $m\ge2$ exist with total cost $O(N^{-1})$.
\end{lemma}
\begin{proof}
Use a uniform mesh of spacing at most $h$, respecting regularity breaks. On each slice set $M_i=X_i(t_{j+1})-X_i(t_j)$ and apply the adjacent transfers successively. Their net change in coordinate $i$ is $M_i-M_{i-1}$, with $M_0=M_d=0$, so each sampled endpoint is exact. The intermediate states are $O(h)$ from $x(t_j)$.

For $M_i\ne0$, absolute continuity supplies a point in the interval where $\dot X_i\ne0$. Bounded derivatives and a positive minimum coordinate make the actual incoming pair ratio at most $R_0+O(h)$. The control in Eq.~\eqref{eq:cyclecontrol} changes it by another $O(h)$, preserving the cap for sufficiently small $h$. All constants are uniform. There are at most $(d-1)O(h^{-1})$ operations, each costing $O(h^2)$. Identity collisions pad unused counts, giving $O(N^{-1})$ with exact endpoints and no correction auxiliary.
\end{proof}

\subsection{Connecting a common envelope class}
Suppose $\B_R(p)=\B_R(q)=r$. The maximal constant-slope blocks of $r_i/\gamma_i$ have strictly decreasing slopes and contact boundaries for both endpoints. Each block thus has the same total mass for $p,q,r$.

For $0<\delta<1$, mix each nonsingleton block of $p$ and $q$ slightly with the uniform distribution of that block, preserving its mass. Denote the resulting states by $p^\delta,q^\delta$. Every proper within-block prefix is now strictly below its geometric envelope prefix. Sorting across blocks persists for small $\delta$: within-block majorization bounds the last upper-block entry from below and the first lower-block entry from above by their respective envelope entries, whose boundary ratio is strictly greater than $R$.

Choose $1<R_0<R$ sufficiently close to $R$ that the geometric profiles of ratio $R_0$, with the same block masses, still majorize $p^\delta$ and $q^\delta$ within every block. Call their concatenation $r^0$. The strict inequalities justify this choice. On a nonsingleton block consider
\begin{equation}
 J_i(x)=[x_{i+1}-x_i/R_0]_+,\qquad
 \dot x_i=J_i-J_{i-1},\label{eq:flow}
\end{equation}
with zero flux at the block boundaries. In cumulative coordinates,
\begin{equation}
 \dot X_i=[X_{i+1}+X_{i-1}/R_0-(1+1/R_0)X_i]_+.
\end{equation}
This is a Lipschitz cooperative system. The cumulatives of $r^0$ are an equilibrium upper bound; comparison preserves within-block majorization by $r^0$. Sorting is invariant: on a face $x_i=x_{i+1}$, both neighboring fluxes are at most $J_i$, so $\dot x_i-\dot x_{i+1}\ge0$. Majorization and sorting imply $x_{\min}\ge r^0_{\min}>0$. They also preserve ordering between blocks.

Each cumulative is nondecreasing and bounded. The nonnegative fluxes have finite integrals and are uniformly continuous on the compact invariant set, so they tend to zero. The limiting block is $R_0$-separated. Every normalized separated block majorizes its geometric profile, whereas the trajectory remains majorized by that profile. The limit is therefore $r^0$.

Run the flows from $p^\delta$ and $q^\delta$ for finite times until both lie close to $r^0$. Join them by a short straight segment preserving block masses and traverse the second flow backward. The bridge has all relevant ratios below $R$. A nonzero flow requires ratio below $R_0$. The complete finite path therefore satisfies Lemma~\ref{lem:sampling} with a common strict margin, enlarged slightly for the bridge if necessary. Its endpoints are exactly $p^\delta,q^\delta$ and its cost tends to zero as the mesh is refined.

The initial regularization $p\to p^\delta$ is available with any fixed $m\ge2$. A uniform auxiliary controls equal-angle two-level rotations with phases $2\pi a/m$, $a=0,\ldots,m-1$. Averaging cancels output pair coherences and realizes a bistochastic two-level mixing. The auxiliary marginal stays uniform, its heat is zero, and the cost is exactly the system's entropy increase. A finite product of pair mixings produces $p^\delta$, costing $S(p^\delta)-S(p)\to0$.

For fixed $\delta$, refine the sampled path; then let $\delta\downarrow0$ along a diagonal sequence. The output approaches $q$ and total cost vanishes. If the original endpoints already have strict proper internal prefixes in every maximal envelope block, set $\delta=0$. Lemma~\ref{lem:sampling} then gives exact endpoints with $O(N^{-1})$ cost. This is the strict-slack result used for product memory return.

\section{Completion of the exact-law proof}\label{app:attainment}
\subsection{Physical middle stages}
For $a=\B_R(p)$ and $b=\B_R(q)\prec a$, separation is preserved along $x(t)=(1-t)a+tb$. The downward slice transfers in Eq.~\eqref{eq:flows} leave every actual state within $O(n^{-1})$ of that path. Its minimum population is positive. Consequently, for fixed $m$,
\begin{equation}
 t_* =\frac{A-(m-1)M}{B+M}\ge R-C/n>1
\end{equation}
for sufficiently large $n$, with a uniform constant $C$. When $t=t_*<R$, Eq.~\eqref{eq:angle} equals one. When $t=R\le t_*$, it lies in $(0,1]$. Thus every transfer is a valid joint rotation.

The auxiliary marginal gains $M$ in level zero and loses $M$ in level one. Since their population ratio is $t$, its energy change is $\beta Q=-M\log t$. Joint off-diagonal entries created by the rotation vanish under both partial traces. This proves Eq.~\eqref{eq:downcost} without a perturbative approximation.

If $t<R$, the estimate $R-t=O(n^{-1})$ and $M=O(n^{-1})$ give $M\log(R/t)=O(n^{-2})$. There are at most $(d-1)n$ such terms. Entropies telescope exactly and $\log\gamma_i=(d-i)\log R-\log Z$, proving Eq.~\eqref{eq:middlecost}. The correction is nonnegative. This argument does not assume that actual intermediate states remain separated.

\subsection{Joining approximate outer stages}
The cost between $p$ and $a$, and between $b$ and $q$, has infimum zero by Appendix~\ref{app:zero}. To combine them rigorously, fix output-error and cost tolerances. First choose a finite middle protocol with sufficiently small excess over Eq.~\eqref{eq:middlecost}, and a finite $b\to q$ protocol with sufficiently small cost and target error. Freeze every auxiliary state and unitary in these two finite protocols. Their composite output and total cost are continuous in the input near $a$, since the states and auxiliaries are full rank. Choose a finite $p\to a$ protocol accurate enough that its perturbation changes the subsequent cost and output by less than the remaining tolerances, and with its own cost below tolerance. Let the tolerances tend to zero.

This gives the upper bound for every $q\prec\B_R(p)$. Conversely, the invariant majorization set $\C_{\B_R(p)}$ contains every attainable output and is closed, so targets outside it cannot even be approached. The lower bound follows from Eqs.~\eqref{eq:telescoping} and \eqref{eq:projectionphi}. Theorem~\ref{thm:law} follows. No uniform continuity constant for an infinite family of increasingly long protocols has been assumed.

At $R=1$ all fresh auxiliaries are uniform. A separate unital argument gives reachability $q\prec p$ and cost $H(q)-H(p)$. Nonidentical ordered spectra then cannot be interconverted at zero cost. Our strict-slack constructions assume $R>1$.

\section{Memory identities and budgets}\label{app:memory}
For the sorted repeated spectrum $p\otimes u_K$, the cumulative probability at group boundary $jK$ is $P_j$. The geometric abscissae obey
\begin{equation}
 \Gamma^{dK,R}_{jK}=\frac{1-R^{-jK}}{1-R^{-dK}}=\Gamma^{d,R^K}_j.
\end{equation}
Inside each group, the probability prefix is linear in its index and the geometric prefix is concave. The point lies below the chord joining its group boundaries in the geometric-abscissa plot. The full concave majorant is therefore determined by coarse boundaries alone. Subdividing the coarse increment geometrically proves Eq.~\eqref{eq:uniform}; the product is sorted because adjacent coarse ratios are at least $R^K$.

For any $R^K$-separated $b$, group-boundary sums and the same chord argument give
\begin{equation}
 q\otimes u_K\prec b\otimes\gamma_{K,R}
 \quad\Longleftrightarrow\quad q\prec b.
\end{equation}
Together with factorization of $\gamma_{dK,R}$, this proves Theorem~\ref{thm:uniform}. For general $c\prec\gamma_{K,T}$, put $b=\B_{T^K}(p)$. Then $p\otimes c\prec b\otimes\gamma_{K,T}$, and the latter is $T$-separated, hence $R$-separated. Minimality implies $\B_R(p\otimes c)\prec b\otimes\gamma_{K,T}$. Applying a $kK$-prefix to a reachable $q\otimes c$ proves Eq.~\eqref{eq:arbitrarymemory}. Enlarging the operational cap to $T$ and using Theorem~\ref{thm:preparable} proves Eq.~\eqref{eq:arbitrarycost}.

For sufficiently large fixed uniform memory, let $s=R^K$ and $s^{-1}<\eps$, where $\eps$ is the smallest endpoint eigenvalue. The tail of the geometric distribution after any $j<d$ satisfies
\begin{equation}
 1-\Gamma^{d,s}_j=\frac{s^{-j}-s^{-d}}{1-s^{-d}}
 \le s^{-j}\le s^{-1}<\eps.
\end{equation}
Every endpoint tail is at least $\eps$, so all proper coarse majorization inequalities are strict. Interior repeated-spectrum points also lie strictly below their geometric chords. The exact-endpoint part of Appendix~\ref{app:zero} therefore proves Proposition~\ref{prop:exact} for any fixed $m\ge2$.

For $T\ge R$, the vector $\B_T(p)$ is an $R$-separated majorant, whence $\B_R(p)\prec\B_T(p)$. Monotonicity in the argument gives $\B_T(\B_R(p))\prec\B_T(p)$; the reverse inequality follows from $p\prec\B_R(p)$. This proves the nesting identity used in the integer memory search.

For a qubit, $\B_s(x,1-x)$ equals $\gamma_{2,s}$ when $x\le s/(1+s)$ and the original spectrum otherwise. Distinct qubit envelopes agree exactly when both lie below that ceiling, yielding Eq.~\eqref{eq:qubitthreshold}. For the domain $p_i\ge\eps$, the extremal vector $v=(1-(d-1)\eps,\eps,\ldots,\eps)$ majorizes every member. It is majorized by $\gamma_{d,s}$ exactly when $v_1\le\gamma_{d,s,1}$: the remaining geometric tail is decreasing, so each tail of length $d-k$ is at most $(d-k)/(d-1)$ times the first tail. Hence the first-prefix inequality implies all the others. Comparing $v$ with the uniform state proves necessity of Eq.~\eqref{eq:domain}; strict inequality makes all relevant prefixes strict and proves its exact-return version.

\section{Permutation-only obstruction}\label{app:perm}
Let $x> x_R=R/(1+R)$ and let $q_j$ denote the auxiliary populations. Then $xq_j>(1-x)q_k$ for all $j,k$. Each output row under a product-eigenbasis permutation selects $m$ entries of these two rows. If it selects an entire original row, the ordered system spectrum is unchanged. Otherwise both selected rows are mixed.

Relative to the complete upper row, a mixed row replaces equal positive numbers of upper and lower entries. Each replacement loses at least
\begin{equation}
 xq_{\min}-(1-x)q_{\max}\ge q_{\min}[x-R(1-x)].
\end{equation}
The argument applies to either output row, so it bounds their larger total. The smallest allowed auxiliary population satisfies $1=\sum_jq_j\le[1+(m-1)R]q_{\min}$, establishing Eq.~\eqref{eq:permgap}. A spectrum with one minimum and $m-1$ maximum populations attains the bound by a single exchange.

No subsequent allowed collision can undo this forbidden interval. Above $x_R$, the largest eigenvalue is nonincreasing by the one-cut Ky Fan argument. The interval $[1/2,x_R]$ is invariant under all cap-$R$ collisions because its spectra are majorized by the separated ceiling $(x_R,1-x_R)$. Therefore, after the first nontrivial permutation change, no later permutation can reach a target strictly between its maximum landing point and the original $x$. Pure system rotations do not change this conclusion when permutations are defined in the current product eigenbasis.

For $x=.85$, $R=3$, take a fresh qubit with populations $(.75,.25)$ and rotate $|0,1\rangle$ with $|1,0\rangle$ using $\sin^2\theta=1/2$. Their population difference is $0.1$, so the transfer is $0.05$, giving output $(.80,.20)$. Its exact cost is Eq.~\eqref{eq:floor}. The resulting joint off-diagonal entry has magnitude $0.05$, while each reduced state is diagonal. Allowing external stochastic mixtures changes the comparison and is not covered by the permutation-only obstruction.

\section{Finite witness and collision allocation}\label{app:finite}
For the path in Eq.~\eqref{eq:path}, equal subdivision uses the seven signed stroke sizes
\begin{equation}
 (.175,.035,-.05,-.05,-.02,-.15,-.04)/n.
\end{equation}
The middle three strokes are interleaved at each of $n$ slices. The reverse sequence reverses their order and sign and retunes each auxiliary to the reverse target pair. In the forward direction all active auxiliary ratios are at most $11/4$. The limiting reverse ratio is $11/4+5/(4n)$, so $n\ge5$ suffices for both directions. All joint populations are at least $1/25$ in these admissible schedules.

For the normalized pair, $D\le\chi^2$ gives
\begin{equation}
 \sigma\le M^2\left[\frac1{a+M}+\frac1{b-M}\right]\le50M^2.
\end{equation}
The sum of the seven squared total stroke sizes is $1227/20000$, giving $\Sigma_N\le1227/(400n)=8589/(400N)$. Exact rational propagation verifies each endpoint, and rational logarithm enclosures certify the original $N=56$ and $63$ comparison. The first multiple of seven for which the coarse analytic bound itself lies below the floor is $945$.

For the improved schedule, group the path into five segments. A pure pair segment has local length density
\begin{equation}
 g(t)=|\dot M(t)|\sqrt{x_i(t)^{-1}+x_j(t)^{-1}}.
\end{equation}
For the middle straight segment let $f_i=\sum_{j\le i}(B_j-A_j)$, $i=1,2,3$. A slice still uses three adjacent transfers, giving its effective density per counted collision
\begin{equation}
 g_{\rm mid}(t)=\sqrt{3\sum_{i=1}^3f_i^2
           [x_i(t)^{-1}+x_{i+1}(t)^{-1}]}.
\end{equation}
Let $L_s$ be the integral of the appropriate density along segment $s$. Uniform arclength sampling within each segment and collision allocation proportional to $L_s$ give the restricted leading coefficient
\begin{equation}
 C_{\rm five}=\frac12\left(\sum_{s=1}^5L_s\right)^2
             \simeq1.14928423985.\label{eq:coefficient}
\end{equation}
The corresponding lengths are approximately $0.391268$, $0.188040$, $0.408944$, $0.343815$, and $0.184036$. Further time sharing among the three middle edges can lower the leading coefficient, but no global optimum is claimed.

The finite implementation in the supplement rounds the sampled interpolation parameters to rationals, preserves both end parameters exactly, assigns integer segment counts, and pads unused collisions by identities. Every intermediate joint permutation, auxiliary normalization, positivity condition, and cap is checked with rational arithmetic. Rounded sampling is used only to design each recorded finite protocol; Eq.~\eqref{eq:coefficient} describes the ideal continuous allocation and is evaluated by numerical quadrature. Directed interval evaluation of Eq.~\eqref{eq:paircost} proves that the improved $N=56$ upper cost bound is below the no-memory lower bound, whereas the improved $N=49$ lower cost bound is above it. This establishes two finite comparisons, not a search over all possible shorter protocols.

For the logarithm certificates, write $z=(x-1)/(x+1)$ for rational $x>0$ and use
\begin{equation}
 \log x=2\sum_{j=0}^{L-1}\frac{z^{2j+1}}{2j+1}+r_L,
 \qquad |r_L|\le\frac{2|z|^{2L+1}}{(2L+1)(1-z^2)}.
\end{equation}
The included calculations use $L=100$ and outward integer rounding at 90 decimal places. Signed coefficients are handled by swapping interval endpoints. The rational bounds, not their rounded decimal displays, are the certificates.

\section{Computational checks}\label{app:checks}
The supplement contains a fresh execution record for the retained tests and the new results. The earlier envelope, flow, exact-sampling, memory, and global-correlation checks are rerun; their provenance is retained. Additional tests check the potential identity, physical middle stages, preparable-memory equivalence, arbitrary-memory bounds, and the permutation obstruction. Counts refer to the supplied code and recorded outputs, not to searches reported in the AI-assisted reviews.

\begin{table}[tbp]
\caption{Selected checks executed for this revision. Exact checks concern finite instances; they do not replace the analytical proofs.}\label{tab:checks}
\begin{ruledtabular}
\begin{tabular}{p{0.63\columnwidth}r}
Check family & Count\\\hline
Potential/projection block identities & 256\\
Rational middle-stage protocols & 102\\
Middle-stage joint population updates & 32\,040\\
Arbitrary coherent collision inequalities & 360\\
Preparable-memory scenarios & 320\\
Endpoint envelope identities in those scenarios & 640\\
Arbitrary-memory scenarios & 320\\
Full-matrix tests of the downward primitive & 240\\
Permutation instances ($m=2,3$) & 4\,464\\
Reallocated finite protocols & 22\\
Joint updates in reallocated protocols & 9\,310
\end{tabular}
\end{ruledtabular}
\end{table}

The preparable-memory scenarios use exact rational spectra over system dimensions $2$--$6$, memory dimensions $2$--$5$, and four caps. They verify Eq.~\eqref{eq:memoryequal} exactly and compare potential values at 70-digit precision. The arbitrary-memory bound is tested on all 127 reachable cases among 320 sampled pairs. The biased-memory example, its strict slack, and the impossibility of its preparation with the system returned are exact rational checks.

The permutation tests enumerate all product-basis permutations for six auxiliary spectra at each of $m=2,3$. Independent linear programs optimize over the whole cap-constrained auxiliary simplex for every mixed row selection. Full-matrix primitive tests cover $d=2,3,4,5$ and $m=2,3,5$. These matrix and optimization comparisons use explicit floating-point tolerances.

The reallocated family has 19 cap-admissible recorded protocols and three rejected coarse reverse schedules at $N=35,42,49$. All recorded schedules have exact endpoints; failed cap tests are retained, not described as admissible implementations. Both directions are admissible for every tested $N\ge56$. Costs are evaluated at 70-digit precision. Separate rational certificates enclose the forward $N=49,56,70$ costs and the no-memory floor.

The inherited test suites include exact rational envelope and projection checks, independent prefix linear programs, coherent sequences, numerical flow convergence, and exact memory-return paths. Numerical integration tests a construction but is not a convergence proof. No proof-assistant formalization, experimental validation, hardware noise threshold, or independent specialist certification is asserted by these computations.

\end{document}